\documentclass[fleqn,usenatbib]{mnras}

\usepackage{newtxtext,newtxmath}

\usepackage[T1]{fontenc}

\DeclareRobustCommand{\VAN}[3]{#2}
\let\VANthebibliography\thebibliography
\def\thebibliography{\DeclareRobustCommand{\VAN}[3]{##3}\VANthebibliography}

\usepackage{graphicx}	
\usepackage{amsmath}	
\usepackage{physics}
\usepackage{subfigure}
\usepackage{chemformula}

\title[Constraining fundamental constants using FRBs]{Constraining Fundamental Constants with Fast Radio Bursts: Unveiling the Role of Energy Scale}

\author[Kalita]{
Surajit Kalita\thanks{E-mail: surajit.kalita@uct.ac.za}
\\
High Energy Physics, Cosmology and Astrophysics Theory (HEPCAT) Group, \\ Department of Mathematics and Applied Mathematics, University of Cape Town, Cape Town 7700, South Africa\\
}

\date{Accepted XXX. Received YYY; in original form ZZZ}

\pubyear{2024}

\begin{document}
\label{firstpage}
\pagerange{\pageref{firstpage}--\pageref{lastpage}}
\maketitle

\begin{abstract}
Understanding physical mechanisms relies on the accurate determination of fundamental constants, although inherent limitations in experimental techniques introduce uncertainties into these measurements. This paper explores the uncertainties associated with measuring the fine-structure constant ($\alpha$) and the proton-to-electron mass ratio ($\mu$) using observed fast radio bursts~(FRBs). We select 50 localized FRBs to quantify the effects of varying this fundamental coupling on the relation between dispersion measure and redshift. By leveraging independent measurements of dispersion measures and redshifts of these FRBs, we constrain the uncertainties in $\alpha$ and $\mu$ approximately to $\Delta\alpha/\alpha=1.99\times 10^{-5}$ and $\Delta\mu/\mu=-1.00\times 10^{-5}$ within the standard $\Lambda$CDM cosmological framework. Remarkably, these constraints improve nearly an order-of-magnitude when considering a dynamical dark energy model. This investigation not only yields one of the most stringent constraints on $\alpha$ and $\mu$ to date but also emphasizes the criticality of accounting for the energy scale of the system when formulating constraints on fundamental parameters.
\end{abstract}




\section{Introduction}

Fast radio bursts (FRBs) are short-duration (millisecond timescale) bright transient phenomena. These events manifest within the radio frequency spectrum, spanning from approximately $100\rm\,MHz$ to $8\rm\,GHz$. Since the initial detection of the first FRB in 2007 by \cite{2007Sci...318..777L}, nearly 700 additional FRBs have been cataloged, with the majority of detections attributed to the Canadian Hydrogen Intensity Mapping Experiment~(CHIME) telescope\footnote{\url{https://www.chime-frb.ca/catalog}}. While a significant portion of these FRBs appear as non-repeating one-off bursts, a subpopulation exhibit repeating behavior, albeit without apparent periodicity. Notably, only one FRB~(FRB\,20200428) has been conclusively traced to an origin within our own Milky Way galaxy, specifically associated with the Galactic soft gamma repeater SGR\,$1935+2154$~\citep{2020PASP..132c4202B,2020Natur.587...59B,2020Natur.587...54C}, while some others have been localized within their respective host galaxies. While various theoretical models invoking compact objects such as white dwarfs~(WDs), neutron stars, or black holes, have been proposed to elucidate the characteristics of FRBs~\citep{2019PhR...821....1P,2020Natur.587...45Z}, the lack of coincident electromagnetic counterparts hinders the definitive identification of a single progenitor mechanism. Addressing this challenge, recent work by \cite{2023MNRAS.520.3742K} suggests that the detection of continuous gravitational waves from FRB sites could help to constrain potential progenitor models, offering a promising avenue for future research.

FRBs, distinguished by their notable characteristics including high flux, very short pulse widths, large dispersion measures (DMs), and their capability to probe the intergalactic medium on cosmological scales, have emerged as valuable tools for investigating various astrophysical and cosmological studies. Leveraging FRB\,150418, \cite{2016PhLB..757..548B} constrained the photon mass to $m_\gamma<1.8\times 10^{-14}\,\rm eV\,c^{-2}$, a limit which strengthens with the inclusion of numerous other FRBs~\citep{2021PhLB..82036596W,2023MNRAS.520.1324L}. Moreover, different research groups have recently employed localized FRBs to estimate the Hubble constant~\citep{2022MNRAS.511..662H,2022MNRAS.515L...1W,2022MNRAS.516.4862J}. Additionally, utilizing data from 12 localized FRBs, \cite{2023MNRAS.523.6264R} established the most stringent constraint on the parameterized post-Newtonian parameter associated with the weak equivalence principle. Furthermore, employing the phenomenon of gravitational lensing in FRBs, \cite{2016PhRvL.117i1301M} derived constraints on the fraction of dark matter attributed to primordial black holes, which were subsequently refined by considering various factors such as FRB microstructures~\citep{2020ApJ...900..122S}, extended mass functions~\citep{2020PhRvD.102b3016L}, plasma lensing~\citep{2022PhRvD.106d3017L}, and enhanced data accuracy~\citep{2020ApJ...896L..11L,2023JCAP...11..059K}.

One of the key objectives in physics is to establish the universality of physical laws by testing fundamental couplings. Among them, the fine-structure constant, denoted by $\alpha$, and the proton-to-electron mass ratio, denoted by $\mu$ are of particular interest. These parameters hold significance as they delineate the strength of the electromagnetic interaction between elementary charged particles. Moreover, being dimensionless quantities, $\alpha$ and $\mu$ are independent of any specific system of units, rendering them fundamentally invariant across reference frames.

A multitude of observational and experimental studies have sought to refine the values of $\alpha$ and $\mu$, which primarily use high-resolution spectroscopy of quasars, powerful distant galaxies emitting intense light. Analyses of absorption spectra from quasar HS\,$1549+1919$ yielded a constraint of $\Delta\alpha/\alpha = (-5.4\pm 3.3_\mathrm{stat} \pm 1.5_\mathrm{sys})\times 10^{-6}$~\citep{Evans:2014yva}, where the subscript terms denote statistical and systematic uncertainties. Similar studies on other quasars like HE\,$0515-4414$ gives yielded a different value~\citep{Kotus:2016xxb}, highlighting the ongoing refinement process. Similar strategies are employed to constrain $\mu$. Employing molecular hydrogen transitions in the quasar  spectra, \cite{king2011new} investigated molecular hydrogen transitions in quasar Q$0528-250$ at a high redshift ($z = 2.811$) and estimated $\Delta \mu /\mu = (0.3 \pm 5.1)\times 10^{-6}$. Subsequent work by \cite{2014PhRvL.113l3002B} on quasar PKS\,$1830-211$ at a lower redshift ($z=0.89$) provided a more stringent constraint $\Delta \mu /\mu = (0.0 \pm 1.0) \times 10^{-7}$. Studies at even higher redshifts ($2 \le z \le 3$) by \cite{Le:2019ijj} focused on 27 [Fe II] emission lines in quasars, establishing $\Delta \mu /\mu < 10^{-5}$. However, observations of 21-cm and molecular hydrogen absorption spectra from systems like J$1337+3152$ at $z \approx 3.17$ slightly relaxed this bound, yielding $\Delta \mu /\mu = (-1.7 \pm 1.7) \times 10^{-6}$~\citep{Srianand:2010un}. Very recently, utilizing 110000 galaxy data from the DESI survey, \cite{2024arXiv240403123J} found $\Delta\alpha/\alpha = (2-3)\times 10^{-5}$. A comprehensive review of these efforts to constrain fundamental constants was presented by~\cite{2017RPPh...80l6902M}.

While cosmological observations provide aforementioned constraints on $\alpha$ and $\mu$, astronomical objects, particularly WDs, offer complementary avenues for such studies. Spectroscopic analysis of specific elements like Fe\,V in WD G191$-$B2B reveals $\Delta \alpha / \alpha = (6.36\pm2.27)\times10^{-5}$~\citep{2021MNRAS.500.1466H}. Similarly, Lyman transitions of molecular hydrogen (\ch{H2}) in the spectra of WDs GD\,133 and G$29-38$ yield the bounds $\Delta \mu /\mu = (-2.7 \pm 4.9) \times 10^{-5}$ and $\Delta \mu /\mu = (-5.8 \pm 4.1) \times 10^{-5}$, respectively~\citep{2014PhRvL.113l3002B}. These findings highlight the potential of WDs for constraining specific fundamental constants through surface spectroscopic measurements~\citep{Berengut:2013dta,2014PhRvL.113l3002B}. Furthermore, through a comparison of the WD mass--radius relation with a simulated dataset of 100 WDs, \cite{2017PhRvD..96h3012M} estimated $\Delta \alpha / \alpha = (2.7 \pm 9.1) \times 10^{-5}$, which was later refined using realistic WD data samples from the {\it Gaia} surveys~\citep{2023ApJ...949...62K,2024MNRAS.527..232U}. These explorations partiallty underscores the significance of the system energy scale when constraining fundamental parameters.

In this article, we investigate the constraints on the parameters $\alpha$ and $\mu$ using observations of localized FRBs. We outline the structure of the article as follows. In Section~\ref{Sec2}, we explore the relationship between the mean DM as a function of source redshift and its dependency on the fluctuation of $\alpha$. In Section~\ref{Sec3}, we detail our data sample consisting of 50 localized FRBs. This data is subsequently employed to calculate the constraint on $\alpha$ under standard $\Lambda$CDM cosmology. We further extend this analysis by encompassing a cosmological model that incorporates dynamical dark energy. In Section~\ref{Sec4}, we proceed to calculate the constraint on $\mu$ and subsequently followed by a comprehensive discussion of our results. Finally, we present our concluding remarks in Section~\ref{Sec5}.

\section{Effect of variation in fine-structure constant on redshift--dispersion measure relation}\label{Sec2}

One of the pivotal characteristics exhibited by FRBs is the dispersion sweep discernible in the frequency-time domain. This phenomenon arises due to the presence of intervening ionized plasma along the propagation path of the radio waves from the source to Earth. The extent of this dispersion is quantified by DM, which reflects the total column density of free electrons encountered by the signal. DM integrates contributions from various regions, including the Milky Way~(MW) Galaxy, its circumgalactic halo, the intergalactic medium~(IGM), and the host galaxy of the FRB. Mathematically it can be represented as
\begin{align}\label{Eq: DM}
    \mathrm{DM} = \mathrm{DM}_\mathrm{MW} + \mathrm{DM}_\mathrm{Halo} + \mathrm{DM}_\mathrm{IGM}(z_\mathrm{S}) + \frac{\mathrm{DM}_\mathrm{Host}}{1+z_\mathrm{S}},
\end{align}
where $z_\mathrm{S}$ is the source redshift, while $\mathrm{DM}_\mathrm{MW}$, $\mathrm{DM}_\mathrm{Halo}$, $\mathrm{DM}_\mathrm{IGM}$, and $\mathrm{DM}_\mathrm{Host}$ are DM contributions from our Galaxy, halo, IGM, and host galaxy, respectively. 

Owing to our reasonable understanding of the Galactic distribution of free electrons, $\mathrm{DM}_\mathrm{MW}$ is fairly modeled. Additionally, \cite{2019MNRAS.485..648P} estimated that the Galactic halo contributes $\mathrm{DM}_\mathrm{Halo}\approx 50-80\rm\,pc\,cm^{-3}$ independent of any contribution from the Galactic interstellar medium. The remaining components $\mathrm{DM}_\mathrm{IGM}$ and $\mathrm{DM}_\mathrm{Host}$ are largely unknown due to the challenges associated with their measurements. Nonetheless, the IllustrisTNG simulation shows that for non-repeating FRBs, the median turns out to be $\mathrm{DM}_\mathrm{Host} = 33(1+z_\mathrm{S})^{0.84}\rm\,pc\,cm^{-3}$, whereas for repeating FRBs, it alters to $35(1+z_\mathrm{S})^{1.08}\rm\,pc\,cm^{-3}$ or $96(1+z_\mathrm{S})^{0.83}\rm\,pc\,cm^{-3}$ depending whether the host is a dwarf galaxy or a spiral galaxy, respectively~\citep{2020ApJ...900..170Z}.

The average $\mathrm{DM}_\mathrm{IGM}$ is given by \cite{2020Natur.581..391M} as
\begin{align}\label{Eq: Macquart}
    \langle \mathrm{DM}_\mathrm{IGM}(z_\mathrm{S})\rangle &= \frac{3c \Omega_\mathrm{b} H_0^2}{8\pi G m_\mathrm{p}} \int_{0}^{z_\mathrm{S}} \frac{f_\mathrm{IGM}(z)\chi(z)(1 + z)}{H(z)} \dd{z},
\end{align}
where $c$ is the speed of light, $G$ is the Newton gravitational constant, $H_0$ is the Hubble constant, $\Omega_\mathrm{b}$ is the baryonic matter density, $m_\mathrm{p}$ is the proton mass, $f_\mathrm{IGM}$ is the baryon mass fraction in the IGM, and $\chi(z)$ is the ionization fraction along the line of sight, given by
\begin{align}
    \chi(z) = Y_\mathrm{H}\chi_{\mathrm{e},\mathrm{H}}(z) + \frac{1}{2}Y_\mathrm{p}\chi_{\mathrm{e},\mathrm{He}}(z), 
\end{align}
with $\chi_{\mathrm{e},\mathrm{H}}$ and $\chi_{\mathrm{e},\mathrm{He}}$ respectively being the ionization fractions of the intergalactic hydrogen and helium, and $Y_\mathrm{H} = 3/4$, $Y_\mathrm{p} = 1/4$ their respective mass fractions. In our calculations, we consider $f_\mathrm{IGM}=0.85$ following \cite{2022ApJ...931...88C}. The Hubble function $H(z)$ encodes the information of underlying cosmology and for the $\Lambda$ cold dark matter~($\Lambda$CDM) formalism, neglecting the contributions from radiation and curvature, it is given by $H(z) = H_0\sqrt{\Omega_\mathrm{m} \left(1+z\right)^3 + \Omega_\Lambda}$ with $\Omega_\mathrm{m}$ and $\Omega_\Lambda$ respectively being the present matter and vacuum density fraction such that $\Omega_\mathrm{m}+\Omega_\Lambda=1$.

As outlined in the Introduction, the objective of this study is to impose constraints on $\alpha$. Let us first define dimensionless couplings $\alpha_\mathrm{p}=Gm_\mathrm{p}^2/\hbar c$ and $\alpha_\mathrm{e}=Gm_\mathrm{e}^2/\hbar c$. Considering that the Planck mass is fixed while allowing for variations in the quantum chromodynamics~(QCD) scale and particle masses,~\cite{Coc:2006sx} obtained the ensuing relationships governing uncertainties in electron and proton masses
\begin{align}\label{Eq: del me}
    \frac{\Delta \alpha_\mathrm{e}}{\alpha_\mathrm{e}} = 2\frac{\Delta m_\mathrm{e}}{m_\mathrm{e}} &= \left(1+\mathsf{S}\right)\frac{\Delta \alpha}{\alpha}
\end{align}
and
\begin{align}\label{Eq: del mp}
    \frac{\Delta \alpha_\mathrm{p}}{\alpha_\mathrm{p}} = 2\frac{\Delta m_\mathrm{p}}{m_\mathrm{p}} &= \left[\frac{8}{5}\mathsf{R} +\frac{2}{5} \left(1+\mathsf{S}\right)\right]\frac{\Delta \alpha}{\alpha},
\end{align}
where $\mathsf{R}$ and $\mathsf{S}$ are dimensionless phenomenological parameters. The values of $\mathsf{R}$ and $\mathsf{S}$ exhibit variability in accordance with observational data. For instance, the Wilkinson Microwave Anisotropy Probe (WMAP) results suggest that $\mathsf{R} \approx 36$ and $\mathsf{S} \approx 160$~\citep{Coc:2006sx}, while a dilaton-type model gives $\mathsf{R} \approx 109$ and $\mathsf{S} \approx 0$~\citep{nakashima2010constraining}. Nonetheless, the values we examine in this research are derived from astrophysical observations of BL Lac object PKS\,$1413+135$ observed at a redshift 0.2467 and frequency around 1200\,MHz) yielding $\mathsf{R}=278\pm24$ and $\mathsf{S}=742\pm65$~\citep{2014MmSAI..85..113M}. Now inserting the above uncertainties in Eq.~\eqref{Eq: Macquart}, we obtain
\begin{align}\label{Eq: Macquart_error}
    \langle \mathrm{DM}_\mathrm{IGM}(z_\mathrm{S})\rangle &= \frac{15}{8\pi}\sqrt{\frac{c}{G\hbar\alpha_\mathrm{p}}}\frac{\Omega_\mathrm{b} H_0^2}{\left[5+(1+4R+S)\frac{\Delta \alpha}{\alpha}\right]} \nonumber\\ &\int_{0}^{z_\mathrm{S}} \frac{f_\mathrm{IGM}(z)\chi(z)(1 + z)}{H(z)} \dd{z}.
\end{align}
It shows the dependency of $\mathrm{DM}_\mathrm{IGM}-z_\mathrm{S}$ relation on the variation in $\alpha$. In our analysis, we compare our data points with this mean DM value to put constraint on $\Delta\alpha/\alpha$.

\section{Constraining fine-structure constant using FRB data}\label{Sec3}

\begin{table*}
\centering
\caption{List of all localized FRBs (until January 2024). $\mathrm{DM}_\mathrm{MW}$ contribution is calculated based on NE2001 model. FRBs in bold indicate that their $\mathrm{DM}_\mathrm{Host}$ values are reported.}
\label{Table: FRB}
\begin{tabular}{|l|l|l|l|l|l|l|}
\hline
Name & $\mathrm{DM}_\mathrm{obs}$ & $\mathrm{DM}_\mathrm{MW}$ & $z_\mathrm{S}$ & Repeater & Reference \\
& $(\rm pc\,cm^{-3})$ & $(\rm pc\,cm^{-3})$  & & (Y/N) & \\
\hline
\textbf{FRB 20121102} & 557.0 & 188.0 & 0.19273 & Y & \cite{2017ApJ...834L...7T}\\
FRB 20171020 & 114.1 & 38.0 & 0.0086 & N & \cite{2018ApJ...867L..10M,2023PASA...40...29L} \\
\textbf{FRB 20180301} & 522.0 & 152.0 & 0.3304 & Y & \cite{2019MNRAS.486.3636P}\\
\textbf{FRB 20180916} & 349.349 & 200.0 & 0.0337 & Y & \cite{2020Natur.577..190M}\\
\textbf{FRB 20180924} & 361.42 & 40.5 & 0.3214 & N & \cite{2019Sci...365..565B}\\
\textbf{FRB 20181030} & 103.5 & 40.0 & 0.0039 & Y & \cite{2021ApJ...919L..24B}\\
\textbf{FRB 20181112} & 589.27 & 102.0 & 0.4755 & N & \cite{2019Sci...366..231P}\\
FRB 20181220 & 209.4 & 126.0 & 0.02746 & N & \cite{2023arXiv231010018B}\\
FRB 20181223 & 112.5 & 20.0 & 0.03024 & N & \cite{2023arXiv231010018B}\\
\textbf{FRB 20190102} & 363.6 & 57.3 & 0.291 & N & \cite{2020Natur.581..391M}\\
FRB 20190110 & 221.6 & 37.1 & 0.12244 & Y & \cite{2024ApJ...961...99I}\\
FRB 20190303 & 222.4 & 29.0 & 0.064 & Y & \cite{2023ApJ...950..134M}\\
FRB 20190418 & 184.5 & 71.0 & 0.07132 & N & \cite{2023arXiv231010018B}\\
FRB 20190425 & 128.2 & 49.0 & 0.03122 & N & \cite{2023arXiv231010018B}\\
\textbf{FRB 20190520} & 1204.7 & 113 & 0.241 & Y & \cite{2022ApJ...931...87O}\\
\textbf{FRB 20190523} & 760.8 & 37.0 & 0.66 & N & \cite{2019ApJ...886..135P}\\
\textbf{FRB 20190608} & 338.7 & 37.2 & 0.1178 & N & \cite{2020ApJ...901..134S}\\
FRB 20190611 & 321.4 & 57.83 & 0.3778 & N & \cite{2020ApJ...903..152H}\\
\textbf{FRB 20190614} & 959.2 & 83.5 & 0.6 & N & \cite{2020ApJ...899..161L}\\
FRB 20190711 & 593.1 & 56.4 & 0.522 & Y & \cite{2021MNRAS.500.2525K}\\
FRB 20190714 & 504.0 & 39.0 & 0.2365 & N & \cite{2020ApJ...903..152H}\\
\textbf{FRB 20191001} & 506.92 & 44.7 & 0.234 & N & \cite{2022MNRAS.512L...1K}\\
FRB 20191106 & 332.2 & 25.0 & 0.10775 & Y & \cite{2024ApJ...961...99I}\\
FRB 20191228 & 297.5 & 33.0 & 0.2432 & N & \cite{2022AJ....163...69B}\\
FRB 20200223 & 201.8 & 45.6 & 0.0602 & Y & \cite{2024ApJ...961...99I}\\
FRB 20200430 & 380.1 & 27.0 & 0.16 & N & \cite{2020ApJ...903..152H}\\
FRB 20200906 & 577.8 & 36.0 & 0.3688 & N & \cite{2022AJ....163...69B}\\
FRB 20201123 & 433.55 & 251.93 & 0.05 & N & \cite{2022MNRAS.514.1961R}\\
\textbf{FRB 20201124} & 413.52 & 123.2 & 0.098 & Y & \cite{2022MNRAS.513..982R}\\
FRB 20210117 & 730.0 & 34.4 & 0.2145 & N & \cite{2023ApJ...948...67B}\\
FRB 20210320 & 384.8 & 42.0 & 0.2797 & N & \cite{2022MNRAS.516.4862J}\\
FRB 20210405 & 565.17 & 516.1 & 0.066 & N & \cite{2024MNRAS.527.3659D}\\
\textbf{FRB 20210603} & 500.147 & 40.0 & 0.177 & N & \cite{2023arXiv230709502C}\\
FRB 20210807 & 251.9 & 121.2 & 0.12927 & N & \cite{2022MNRAS.516.4862J}\\
FRB 20211127 & 234.83 & 42.5 & 0.0469 & N & \cite{2022MNRAS.516.4862J}\\
FRB 20211212 & 206.0 & 27.1 & 0.0715 & N & \cite{2022MNRAS.516.4862J}\\
FRB 20220207 & 262.38 & 79.3 & 0.04304 & N & \cite{2023arXiv230703344L}\\
FRB 20220307 & 499.27 & 135.7 & 0.248123 & N & \cite{2023arXiv230703344L}\\
FRB 20220310 & 462.24 & 45.4 & 0.477958 & N & \cite{2023arXiv230703344L}\\
FRB 20220319 & 110.95 & 65.25 & 0.011 & N & \cite{2023arXiv230101000R}\\
FRB 20220418 & 623.25 & 37.6 & 0.622 & N & \cite{2023arXiv230703344L}\\
FRB 20220506 & 396.97 & 89.1 & 0.30039 & N & \cite{2023arXiv230703344L}\\
FRB 20220509 & 269.53 & 55.2 & 0.0894 & N & \cite{2023arXiv230703344L}\\
\textbf{FRB 20220610} & 1458.0 & 31.0 & 1.017 & N & \cite{2023arXiv231110815G}\\
FRB 20220825 & 651.24 & 79.7 & 0.241397 & N & \cite{2023arXiv230703344L}\\
FRB 20220912 & 219.46 & 125 & 0.077 & Y & \cite{2023ApJ...949L...3R}\\
FRB 20220914 & 631.28 & 55.2 & 0.1139 & N & \cite{2023arXiv230703344L}\\
FRB 20220920 & 314.99 & 40.3 & 0.158239 & N & \cite{2023arXiv230703344L}\\
FRB 20221012 & 441.08 & 54.4 & 0.284669 & N & \cite{2023arXiv230703344L}\\
FRB 20230718 & 476.6 & 393.0 & 0.0357 & N & \cite{2024ApJ...962L..13G}\\
\hline
\end{tabular}
\end{table*}

\begin{figure}
	\centering
	\includegraphics[scale=0.5]{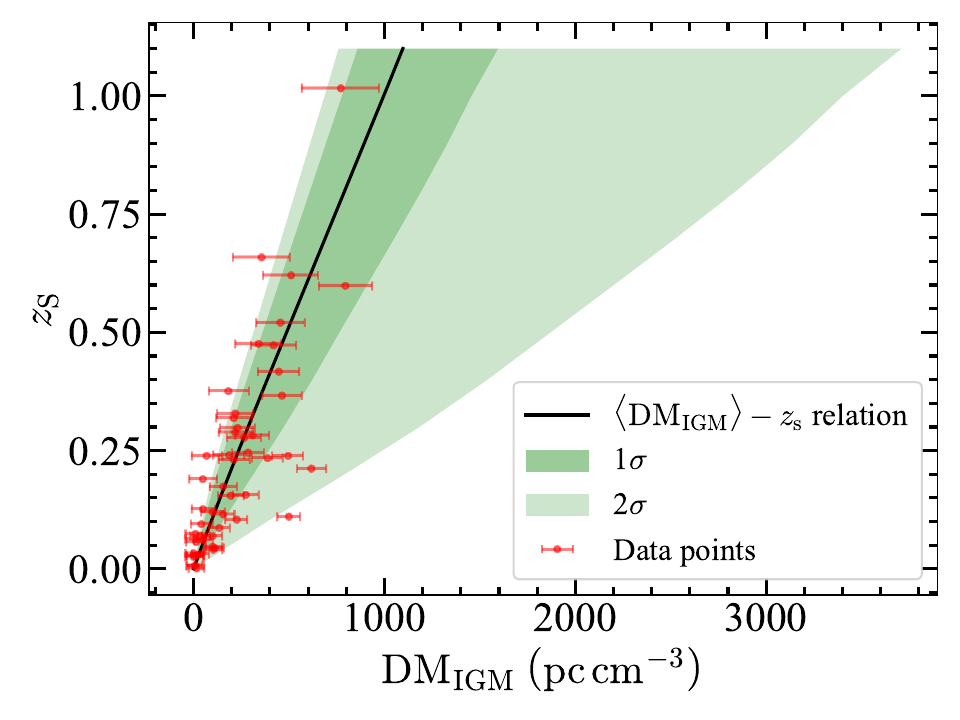}
	\caption{The host redshift values for the localized FRBs are plotted against their estimated DM contribution from IGM along with their error bars. The black solid line depicts the $\langle \mathrm{DM}_\mathrm{IGM}(z_\mathrm{S})\rangle$ as a function of $z_\mathrm{S}$ assuming $\Lambda$CDM cosmology with $H_0=73\rm\,km\,s^{-1}\,Mpc^{-1}$, $\Omega_\mathrm{m} = 0.30966$, $\Omega_\Lambda = 1-\Omega_\mathrm{m}$, and $\Omega_\mathrm{b}=0.04897$. Green shaded regions represent 1$\sigma$ and 2$\sigma$ confidence regions of $\mathrm{DM}_\mathrm{IGM}$.}
	\label{Fig: DM_z}
\end{figure}

In this study, we analyze 50 FRBs that have been accurately localized within their host galaxies as of January 2024. In other words, the source redshift $z_\mathrm{S}$ of these FRBs are measured. The details of these FRBs are given in Table~\ref{Table: FRB}. Utilizing the NE2001 model~\citep{2002astro.ph..7156C}, we compute the Galactic distribution of free electrons and subsequently determine $\mathrm{DM}_\mathrm{MW}$ for each of them. $\mathrm{DM}_\mathrm{Halo}$ is calculated using the model by \cite{2019MNRAS.485..648P} as discussed in the previous section. To account for the contribution of $\mathrm{DM}_\mathrm{Host}$, we divide these 50 FRBs in two categories. For those with measured (reported) $\mathrm{DM}_\mathrm{Host}$, we directly incorporate these values, whereas for FRBs with unknown $\mathrm{DM}_\mathrm{Host}$, we employ the \cite{2020ApJ...900..170Z} model (as described previously) to estimate these missing values. Substituting all these values in Eq.~\eqref{Eq: DM}, we obtain $\mathrm{DM}_\mathrm{IGM}$ for each of these FRBs. Fig.~\ref{Fig: DM_z} shows $z_\mathrm{S}$ values of these FRBs plotted against the calculated $\mathrm{DM}_\mathrm{IGM}$ and their errorbars along with $\langle \mathrm{DM}_\mathrm{IGM}(z_\mathrm{S})\rangle$ assuming standard $\Lambda$CDM cosmology with $H_0 = 73\rm\,km\,s^{-1}\,Mpc^{-1}$, $\Omega_\mathrm{m} = 0.30966$, $\Omega_\Lambda = 1-\Omega_\mathrm{m}$, and $\Omega_\mathrm{b}=0.04897$. It also depicts 1$\sigma$ and 2$\sigma$ confidence regions of $\langle\mathrm{DM}_\mathrm{IGM}\rangle$ following the results obtained by \cite{2019MNRAS.484.1637J} using the Illustris simulation which takes into account for the inhomogeneous distribution of ionized gas in the IGM.

It is important to note that some $\mathrm{DM}_\mathrm{Host}$ values derived from \cite{2020ApJ...900..170Z} simulations appear to be lower in comparison to the currently available data, which indicate significantly larger host DMs~\citep{2024arXiv240308611T}. Hence $\mathrm{DM}_\mathrm{IGM}$ is attributed to a probability distribution, given by \cite{2020Natur.581..391M} as
\begin{align}
    P_\mathrm{IGM}\left(\Delta_\mathrm{IGM}\right) = A \Delta_\mathrm{IGM}^{-\beta_2} \exp[-\frac{\left(\Delta_\mathrm{IGM}^{-\beta_1}-C_0\right)^2}{2\beta_1^2\sigma_\mathrm{DM}^2}],
\end{align}
where $\Delta_\mathrm{IGM} = \mathrm{DM}_\mathrm{IGM}/\langle\mathrm{DM}_\mathrm{IGM}\rangle$, $\sigma_\mathrm{DM}$ is its standard deviation, $A$, $\beta_1$, $\beta_2$, and $C_0$ are model parameters. The values of these parameters are chosen from \cite{2021ApJ...906...49Z} which is based on IllustrisTNG simulation. Similarly, the inherent difficulty in measuring $\mathrm{DM}_\mathrm{Host}$, introduce significant uncertainty regarding its exact distribution to characterize its statistical properties. Thus following \cite{2020Natur.581..391M}, we adopt the following log-normal probability distribution
\begin{align}
    P_\mathrm{Host}\left(\mathrm{DM}_\mathrm{Host}\right) = \frac{1}{\sqrt{2\pi}\mathrm{DM}_\mathrm{Host}\sigma_\mathrm{Host}} \exp[-\frac{\left(\ln{\mathrm{DM}_\mathrm{Host}}-\mu_\mathrm{Host}\right)^2}{2\sigma_\mathrm{Host}^2}],
\end{align}
with $e^{\mu_\mathrm{Host}}$ being the median and $\left(e^{\sigma_\mathrm{Host}^2}-1\right)e^{2\mu_\mathrm{Host}+\sigma_\mathrm{Host}^2}$ its variance. In our analysis, we choose values of $\mu_\mathrm{Host}$ and $\sigma_\mathrm{Host}$ from \cite{2020Natur.581..391M}. These distribution functions effectively capture the inherent uncertainties associated with $\mathrm{DM}_\mathrm{IGM}$ and $\mathrm{DM}_\mathrm{Host}$ values. It is also worth noting that due to precise localization of these FRBs, there is minimal uncertainty associated with their redshift values.

\begin{figure}
	\centering
	\includegraphics[scale=0.5]{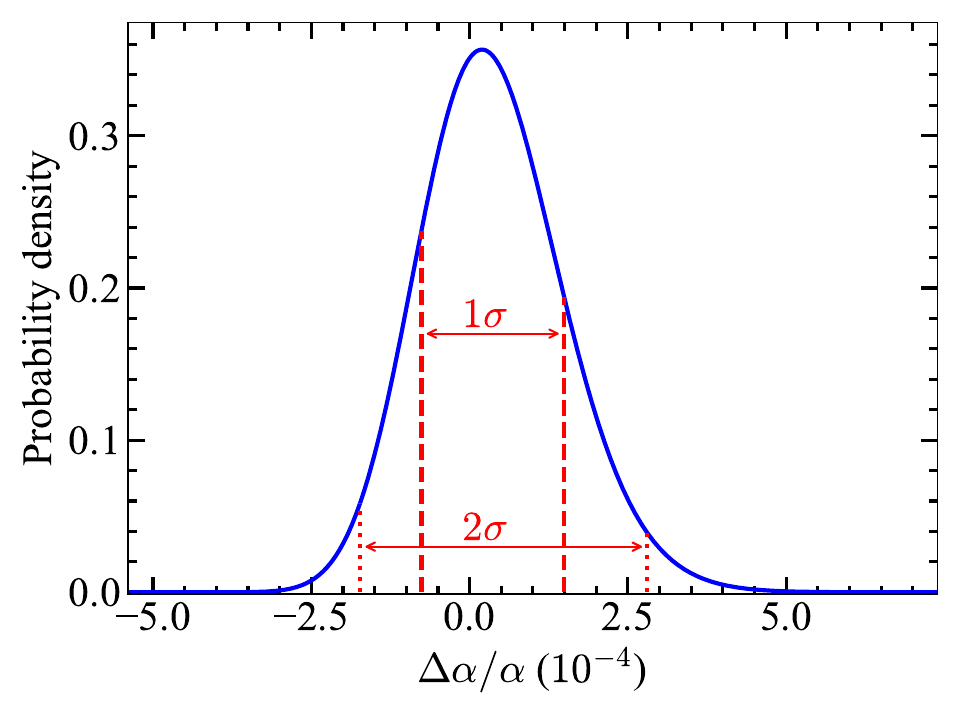}
	\caption{Probability distribution of joint likelihood function with respect to $\Delta\alpha/\alpha$ along with its 1$\sigma$ and 2$\sigma$ confidence intervals.}
	\label{Fig: Likelihood}
\end{figure}

Our objective is to constrain $\Delta\alpha/\alpha$ using this data sample, necessitating an assessment of the theoretical fidelity of the $\langle\mathrm{DM}_\mathrm{IGM}\rangle-z_\mathrm{S}$ curve over these data points. Consequently, we define the following joint likelihood function~\citep{2020Natur.581..391M}
\begin{align}\label{Eq: Likelihood}
    \mathcal{L} = \prod_{i=1}^{N_\mathrm{FRB}} P_i\left(\mathrm{DM}_{\mathrm{excess},i}\mid z_{\mathrm{S},i}\right),
\end{align}
which requires maximization with respect to $\Delta\alpha/\alpha$ where
\begin{align}
    P_i\left(\mathrm{DM}_{\mathrm{excess},i}\mid z_{\mathrm{S},i}\right) &= \int_0^{\mathrm{DM}_{\mathrm{excess},i}} P_\mathrm{Host}\left(\mathrm{DM}_\mathrm{Host}\right) \nonumber \\ &\times P_\mathrm{IGM}\left(\mathrm{DM}_{\mathrm{excess},i}-\frac{\mathrm{DM}_\mathrm{Host}}{1+z_{\mathrm{S},i}}\right)\dd{\mathrm{DM}_\mathrm{Host}}.
\end{align}
Here the subscript $i$ denotes individual data points, $N_\mathrm{FRB}$ represents the total count of localized FRBs in the dataset, and
\begin{align}
    \mathrm{DM}_\mathrm{excess} = \mathrm{DM} - \mathrm{DM}_\mathrm{MW} - \mathrm{DM}_\mathrm{Halo} = \mathrm{DM}_\mathrm{IGM}(z_\mathrm{S}) + \frac{\mathrm{DM}_\mathrm{Host}}{1+z_\mathrm{S}}.
\end{align}
Fig.~\ref{Fig: Likelihood} illustrates the probability density function of the joint likelihood of all 50 localized FRBs with respect to $\Delta\alpha/\alpha$ alongside different confidence intervals. This likelihood function is maximized for $\Delta\alpha/\alpha \approx 1.99^{+12.93}_{-9.61}\times 10^{-5}$ within 1$\sigma$ confidence level, thereby establishing the most robust constraint on $\alpha$ derived from the localized FRB dataset within the framework of the $\Lambda$CDM cosmology.

\subsection{Effect of dynamical dark energy equation of state on constraining the fine-structure constant}

The $\Lambda$CDM cosmological model, despite its simplicity and efficacy in encapsulating the Universe, confronts recent challenges. Foremost among these challenges is the Hubble tension, characterized by a disagreement between the ascertained values of $H_0$ obtained from observations of early and late time cosmologies. Measurements of the cosmic microwave background~(CMB) radiation by the Planck satellite suggest a value of approximately $H_0=67.36\pm0.54\rm\,km\,s^{-1}\,Mpc^{-1}$~\citep{2020A&A...641A...6P}, while using local distance indicators like type Ia supernovae and cepheid variable stars yields a higher value of $H_0=73.04\pm1.04\rm\,km\,s^{-1}\,Mpc^{-1}$~\citep{2022ApJ...934L...7R}. This discrepancy has motivated exploration of alternative physics and cosmological models (see comprehensive discussions by \cite{2021CQGra..38o3001D} and \cite{2023ARNPS..73..153K} on these models). 

\begin{figure}
	\centering
	\includegraphics[scale=0.52]{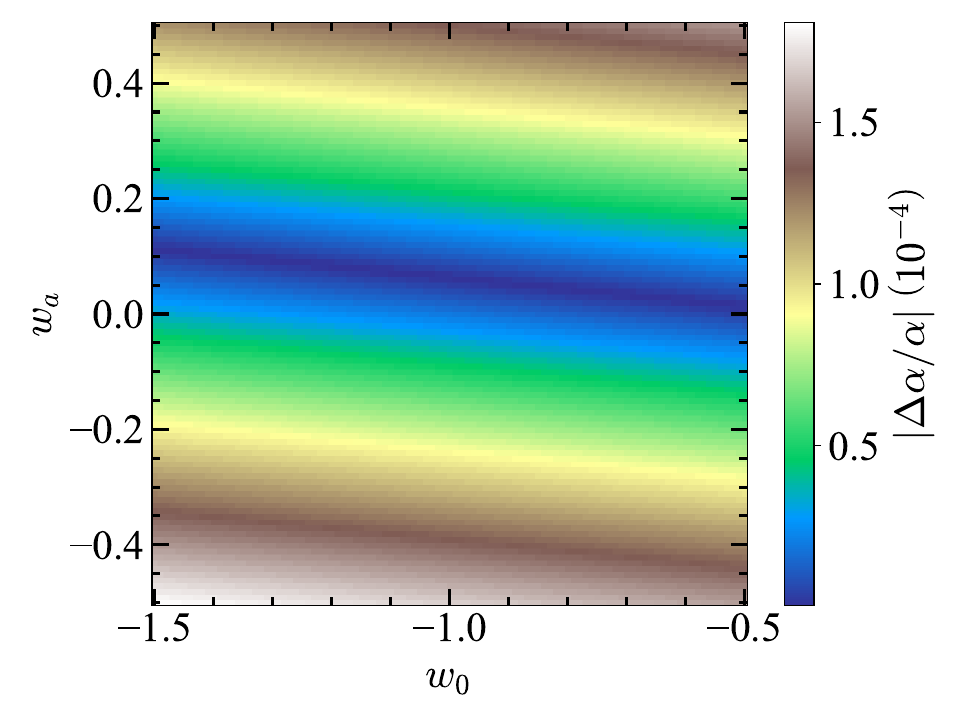}
	\caption{Variation of $\Delta\alpha/\alpha$ for different dark energy equation of states. The colorbar shows the values of $\Delta\alpha/\alpha$ for certain combinations of $w_0$ and $w_a$.}
	\label{Fig: Del_alpha}
\end{figure}

In this study, we consider a simple model with a dynamical dark energy equation of state $w\equiv P/\rho$ with $P$ and $\rho$ respectively being the pressure and energy density. A dynamical dark energy model can potentially reduce both the Hubble tension and the $\sigma_8$ tension simultaneously~\citep{2020PhRvD.101l3516A,2021CQGra..38o3001D,2023PDU....4201266D}, although the specific effects of a free equation-of-state parameter within this framework remain an active area of investigation. This allows for a time-varying Hubble function expressed as $H(z) = H_0\sqrt{\Omega_\mathrm{m} \left(1+z\right)^3 + \Omega_\Lambda f(z)}$ where
\begin{align}
    f(z) = \exp[3\int_0^z \frac{1+w(z')\dd{z'}}{1+z'}].
\end{align}
Assuming the dark energy equation of state varies with time and the function $w(z)$ is parameterized by \cite{2005PhRvD..72j3503J} as
\begin{align}
    w(z) = w_0 + w_a \frac{z}{1+z},
\end{align}
where $w_0$ and $w_a$ are dimensionless parameters. This leads to the following modified form of $f(z)$
\begin{align}
    f(z) = (1+z)^{3\left(1+w_0+w_a\right)} \exp[-3 w_a \frac{z}{1+z}].
\end{align}
Notably $w_0=-1$ and $w_a=0$ recover the standard $\Lambda$CDM cosmology with $w(z)=-1$. However, other parameter combinations can lead to $w(z)<-1$ for certain redshifts. This modification to the cosmological model also impacts $\langle\mathrm{DM}_\mathrm{IGM}(z)\rangle$, which in turn affects the likelihood function of Eq.~\eqref{Eq: Likelihood}. Fig.~\ref{Fig: Del_alpha} illustrates $\Delta\alpha/\alpha$ at which the joint likelihood function is maximized for different combinations of $w_0$ and $w_a$. We observe an improvement in the constraint on $\Delta\alpha/\alpha$ for this dynamical dark energy equations of state. Notably, the most stringent constraint becomes $\Delta\alpha/\alpha \approx 5.4\times10^{-7}$, nearly one order of magnitude lower than the case for $w=-1$.

\section{Discussion}\label{Sec4}

This study investigates the fine-structure constant using DM values of localized FRBs. Analysis is conducted on a sample of 50 FRBs, whose $\mathrm{DM}_\mathrm{IGM}$ is determined by subtracting other contributions from the observed DM. These $\mathrm{DM}_\mathrm{IGM}$ values are then compared to $\langle\mathrm{DM}_\mathrm{IGM}\rangle$, which is theoretically linked to $\Delta\alpha/\alpha$. A likelihood function is further formulated for maximization over $\Delta\alpha/\alpha$, employing parameters derived from the $\Lambda$CDM cosmology. The resulting maximized value is found to be $\Delta\alpha/\alpha \approx 1.99^{+12.93}_{-9.61}\times 10^{-5}$ at the 1$\sigma$ confidence level. Remarkably, this constraint gets a nearly order-of-magnitude improvement while considering a dynamical dark energy model with $w(z)<-1$ for certain redshift. It is worth mentioning that the limitations in our current understanding on the probability distributions of $\mathrm{DM}_\mathrm{IGM}$ and $\mathrm{DM}_\mathrm{Host}$ introduce inherent uncertainties in the derived constraints on the parameter $\alpha$. Consequently, these results should be interpreted with caution until more robust data allows for a more comprehensive characterization of the relevant distributions.

Moreover, using Eqs.~\eqref{Eq: del me} and \eqref{Eq: del mp}, one can obtain the following relation between the uncertainties of the fine-structure constant and proton-to-electron mass ratio
\begin{equation}
    \frac{\Delta \mu}{\mu} = \left[\frac{4}{5}\mathsf{R}-\frac{3}{10}\left(1+\mathsf{S}\right)\right]\frac{\Delta \alpha}{\alpha}.
\end{equation}
Substituting $\Delta\alpha/\alpha = 1.99^{+12.93}_{-9.61}\times 10^{-5}$, we obtain $\Delta\mu/\mu = -1.00^{+3.81}_{-7.47}\times 10^{-5}$, with potential improvement by another order of magnitude for the aforementioned dynamical dark energy model. These constraints on $\alpha$ and $\mu$ outperform many previous reported constraints based on quasar or WD data (mentioned in the Introduction), despite the relatively large error bars attributable to the limited sample size (only 50 FRBs with known redshifts). Anticipated improvements in these error bars are foreseeable with increased data availability, particularly from additional localized FRBs.

Within the context of the $\Lambda$CDM cosmological model, we posit the existence of characteristic physical scales such as the Hubble scale, horizon scale, and baryon acoustic oscillation scale. Elucidating these scales is paramount for investigations into the large-scale structure and evolution of the universe. Deviations from the $\Lambda$CDM formalism (e.g. the aforementioned dynamical dark energy model) necessitate a modification of both the length and energy scales governing the system. This, in turn, entails a restructuring of the characteristic timescales across different cosmological epochs. This study underscores the criticality of considering the inherent scales of a system when establishing constraints on any given physical parameter. Such an approach is anticipated to play an instrumental role in rigorously testing various cosmological models and alternative theories of gravity. It is important to emphasize that we do not advocate for the primacy of any specific dynamical dark energy model. Rather we emphasize the utilization of observations across diverse energy regimes to constrain fundamental parameters.

FRBs offer the advantage of probing a comparatively lower redshift range than CMB or other traditional cosmological observations. Conversely, FRBs are among the rare astronomical phenomena (alongside quasars) detectable at relatively high redshifts, establishing their unique value as a cosmological probe. Furthermore, alteration of dark energy equation-of-state alter underlying physics (e.g., rate of acceleration of the Universe) at identical stages of the Universe's evolution. This inherently translates to alterations in the system's characteristic length scale, which is directly linked to its energy scale. Our findings demonstrate that the constraints imposed on $\alpha$ and $\mu$ from FRB observations exhibit sensitivity to deviations from the $\Lambda$CDM cosmological model. Consequently, the system's energy scale plays a critical role in constraining these fundamental parameters.   

\section{Conclusion}\label{Sec5}

Our work stands out for providing not only some of the tightest constraints on $\alpha$ and $\mu$ using localized FRBs, but also sheds light on how these constraints depend on the specific theory of gravity employed and its characteristic energy scale. Despite the current count of detected FRBs remaining relatively modest, projections estimate an anticipated rate of occurrence ranging from approximately 100 to 1000 events per day~\citep{2016MNRAS.460L..30C}. Ongoing efforts, bolstered by forthcoming observational facilities such as the Hydrogen Intensity and Real-time Analysis eXperiment~(HIRAX), the Deep Synoptic Array~(DSA)-2000, and the Bustling Universe Radio Survey Telescope in Taiwan~(BURSTT), promise a significant enhancement in the detection rates for FRBs in the near future. Consequently, this advancement will facilitate further refinement of the aforementioned constraints.


\section*{Acknowledgments}

The author would like to thank the editor and anonymous reviewer for their constructive comments to improve the manuscript content, particularly with the data analysis portion. Further thanks to Shruti Bhatporia of UCT for her invaluable assistance in selecting the data sample and Amanda Weltman of UCT for having important discussion during the compilation of this work. Acknowledgments are extended to the South African Research Chairs Initiative (SARChI) of the Department of Science and Technology~(DST) and the National Research Foundation~(NRF) for their support. Computations were performed using facilities provided by the University of Cape Town’s ICTS High Performance Computing team: \href{https://ucthpc.uct.ac.za/}{hpc.uct.ac.za}.

\section*{Data availability}
The FRB data used in this article were accessed from \url{https://www.chime-frb.ca/catalog}, \url{https://www.herta-experiment.org/frbstats/catalogue}, and \url{https://frbcat.org}. The derived data generated in this research will be shared on reasonable request to the author.



\bibliographystyle{mnras}
\bibliography{Bibliography} 


\bsp	
\label{lastpage}
\end{document}